\documentclass[a4paper,11pt]{article}
\usepackage{pos}
\usepackage{subcaption}
\usepackage{siunitx}
\usepackage{comment}
\title{Transient studies using a TCAD and Allpix Squared combination approach}
\author*[a,b]{Manuel A. Del Rio Viera}
\author[c]{Eric Buschmann}
\author[a]{Ankur Chauhan}
\author[c]{Dominik Dannheim}
\author[c,e]{Katharina Dort}
\author[a]{Doris Eckstein}
\author[a]{Finn Feindt}
\author[a]{Ingrid-Maria Gregor}
\author[a]{Karsten Hansen}
\author[a]{Yajun He}
\author[a]{Lennart Huth}
\author[a,b]{Larissa Mendes}
\author[a]{Budi Mulyanto}
\author[a,f]{Daniil Rastorguev}
\author[a]{Christian Reckleben}
\author[a,b]{Sara Ruiz Daza}
\author[a,g]{Judith Schlaadt}
\author[a]{Paul Schütze}
\author[a,b]{Adriana Simancas}
\author[c]{Walter Snoeys}
\author[a]{Simon Spannagel}
\author[a]{Marcel Stanitzki}
\author[a]{Anastasiia Velyka}
\author[a,b]{Gianpiero Vignola}
\author[a]{Håkan Wennlöf}

\affiliation[a]{Deutsches Elektronen-Synchrotron DESY, Notkestr. 85, 22607 Hamburg, Germany}
\affiliation[b]{University of Bonn, Regina-Pacis-Weg 3, Bonn, 53113, Germany}
\affiliation[c]{Conseil Européen pour la Recherche Nucléaire, Esplanade des Particules 1, Geneva 23, 1211, Switzerland}
\affiliation[d]{University of Hamburg, Mittelweg 177, Hamburg, 20148, Germany}
\affiliation[e]{University of Giessen, Ludwigstraße 23, Gießen, 35390, Germany}
\affiliation[f]{University of Wuppertal, Gaußstraße 20, Wuppertal, 42119, Germany}
\affiliation[g]{Johannes Gutenberg University of Mainz, Saarstraße 21, Mainz, 55122, Germany}

\emailAdd{manuel.del.rio.viera@desy.de}

\FullConference{%
  6th International Conference on Technology and Instrumentation in Particle Physics (TIPP2023)\\
  4 - 8 Sep 2023\\
  Cape Town, Western Cape, South Africa
}

\abstract{
The goal of the TANGERINE project is to develop the next generation of monolithic silicon pixel detectors using a \SI{65}{\nano\meter} CMOS imaging process, which offers a higher logic density and overall lower power consumption compared to previously used processes. A combination of Technology Computer-Aided Design (TCAD) and Monte Carlo (MC) simulations is used to understand the physical processes within the sensing element and thus the overall performance of the pixel detector. The response of the sensors can then be tested in laboratory and test beam facilities and compared to simulation results.

Transient simulations allow for studying the response of the sensor as a function of time, such as the signal produced after a charged particle passes through the sensor. The study of these signals is important to understand the magnitude and timing of the response from the sensors and improve upon them. While TCAD simulations are accurate, the time required to produce a single pulse is large compared to a combination of MC and TCAD simulations.

In this work, a validation of the transient simulation approach and studies on charge collection are presented. \hfill\break

}

\begin{document}
\maketitle

\section{Introduction}

The TANGERINE project \cite{TANGERINE} aims for the development of the next generation of monolithic silicon pixel sensors and investigates their possible use as vertex detectors in future lepton colliders. Monolithic sensors are designed using a \SI{65}{\nano\meter} CMOS imaging process, which offers a higher logic density and overall lower power consumption compared to currently utilized feature sizes. Vertex detectors in future lepton colliders require a spatial resolution of around \SI{3}{\micro\meter}, good timing performance on the order of nanoseconds, and a low material budget of less than 1\% $X_{0}$ to fulfill its vertexing performance and physics goals. Monolithic Active Pixel Sensors (MAPS) with a small collection electrode are investigated as sensor candidates. A small collection electrode reduces the sensor capacitance and improves the signal-to-noise ratio.

Developing a new sensor is a highly complex process and sensor submissions are time-consuming and costly. Therefore, it is crucial to simulate the sensor response to charged particles as precisely as possible. This work presents a combined approach using Technology Computer-Aided Design (TCAD) and Monte Carlo (MC) simulations which allows to predict the sensor performance and thus can reduce the required number of sensor submissions.

The sensors used for the validation of the approach are the ALICE Analogue Pixel Test Structure (APTS) chips \cite{APTS}, which are produced in the \SI{65}{\nano\meter} CMOS imaging process designed at CERN for the CERN EP R\&D program and are part of the studies for the ALICE ITS3 upgrade \cite{ALICE}. They consist of a 4x4 pixel matrix with analog output and pitches ranging from \SI{10}{\micro\meter} to \SI{25}{\micro\meter} and a thickness of \SI{50}{\micro\meter}.  In this work, two different sensor layouts, Standard and N-Gap \cite{Ngap} with a pixel pitch of \SI{25}{\micro\meter} and a source follower as an output buffer were studied.

To investigate the differences in performance that these designs produce, simulation studies on the charge collection and the timing capabilities of these sensors were performed.

\begin{figure} [h]
\begin{subfigure}{.5\textwidth}
  \centering
  \includegraphics[width=.7\linewidth]{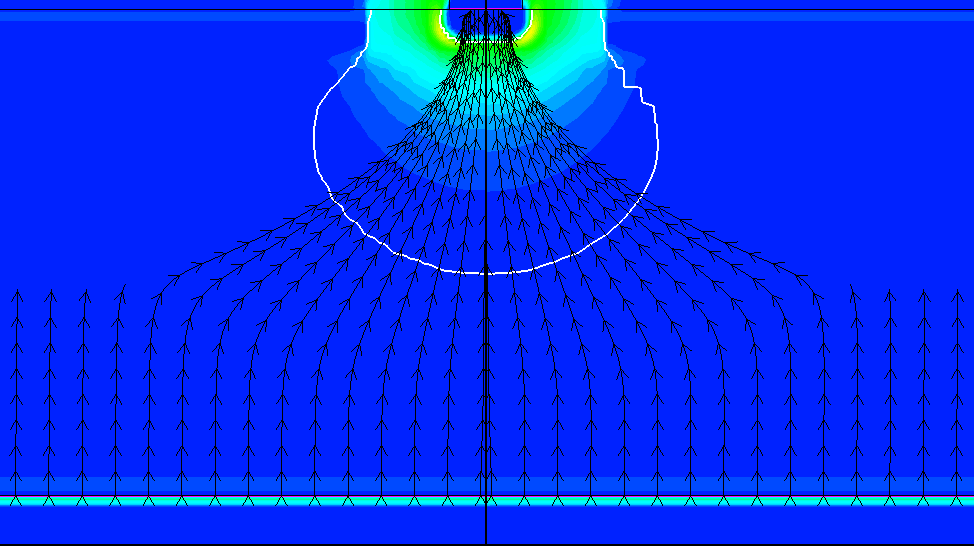}
  \caption{Standard}
  \label{fig:Standard}
\end{subfigure}
\begin{subfigure}{.5\textwidth}
  \centering
  \includegraphics[width=.7\linewidth]{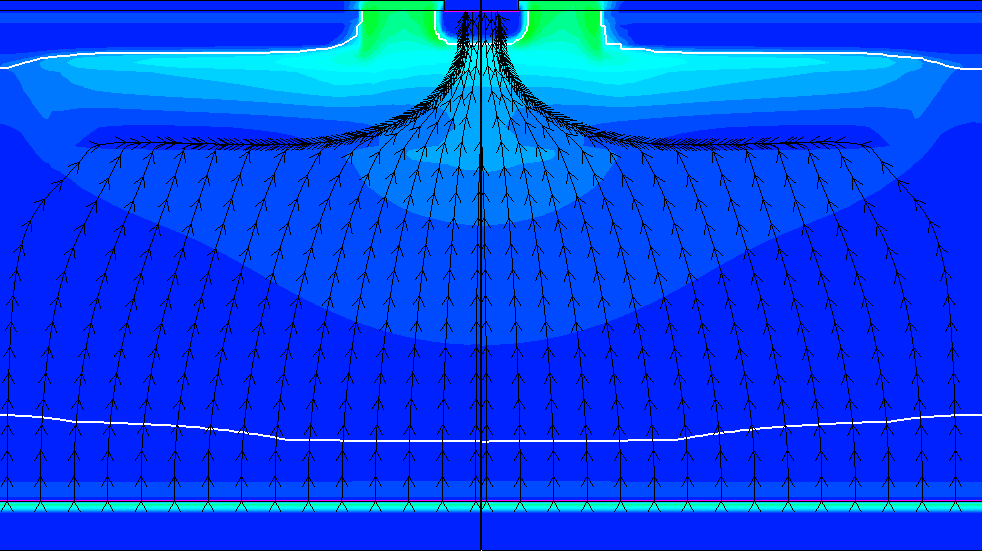}
  \caption{N-Gap}
  \label{fig:Ngap}
\end{subfigure}
\caption{Electric field maps and streamlines of two sensor designs generated via TCAD \cite{TCAD-Adri}. The electrons follow the direction of the lines toward the collection electrode.}
\label{fig:Layouts}
\end{figure}
 
\begin{comment}
\subsection{Schokley-Ramo Theorem}
After a charged particle travels through the sensor, the charge carriers produced after ionization will drift following the electric field toward the collection electrode. The signal induced in the collection electrode is calculated through the Shockley-Ramo theorem \cite{Ramo}, which states that the induced current is proportional to the moving charge $q$, the weighting field $\overrightarrow{E_{w}}$, and the component of the drift velocity $\overrightarrow{\nu}$ parallel to it according to:  
\begin{equation}
    I_{ind} = q \overrightarrow{E_{w}} \cdot \overrightarrow{\nu}
    \label{Ramo}
\end{equation}

The weighting field $\overrightarrow{E_{w}}$ describes the electromagnetic coupling of a charge to an arrangement of conducting electrodes and it is dependent on the geometry of the arrangement.

Often easier to use, the weighting potential $\phi$ appears as a solution to the Laplace equation of the weighting field by integrating equation (\ref{Ramo}):

\begin{equation}
    Q_{ind} (t)= -q [\phi(x(t))-\phi(x_0)]
    \label{Charge_coll}
\end{equation}

$x_{0}$ is the initial position of the moving charge and $x(t)$ is the position of the moving charge at time $t$. Equation (\ref{Charge_coll}) allows to calculate the average charge that is induced at the collection electrode as a function of time known as the collected charge. 
\end{comment}

\subsection{Technology Computer-Aided Design}

To understand better how these different layouts perform in terms of timing and charge collection, it is of vital importance to produce realistic electric fields to ensure precise simulation results.

Technology Computer-Aided Design (TCAD) allows the modeling of semiconductor devices using finite element methods, which enables to produce highly complex electric fields as in the case of thin sensors with a small collection electrode shown in Figure \ref{fig:Layouts}. These are technology-independent simulations, which can only describe the pure sensor capabilities due to layout geometry but do not include any internal well structure. With the use of generic doping profiles, the electric field and electrostatic potential are obtained. 

A simulation of the electric field of these layouts is shown in Figure \ref{fig:Layouts}. In the Standard design (Fig. \ref{fig:Standard}), a depleted region is created around the small collection electrode. Electrons within the depletion region will primarily drift following the direction of the streamlines toward the collection electrode and electrons generated outside this region primarily undergo diffusion. Collection by diffusion is a slow process compared to drift collection. To increase the depleted region and thus reduce the charge collection time, a low-doped layer with a gap is introduced. This creates a lateral component on the electric field indicated by the horizontal arrows shown in Figure \ref{fig:Ngap} and results in a faster charge collection.  

\subsection{The Allpix Squared framework}
While TCAD is an indispensable tool, the time required to simulate and produce results is large for high statistic simulations. Instead, a combination of TCAD and Monte Carlo simulations can be applied to reduce the simulation time drastically.  Allpix Squared (APSQ)  \cite{Allpix} is a simulation framework for semiconductor detectors. It offers a full detector simulation chain and integration with GEANT4 \cite{Geant4} and TCAD. High statistic simulations while maintaining high precision and more realistic scenarios are possible by importing doping concentrations, electric fields, and weighting potentials generated in TCAD into the framework.

\section{Simulation Results}
\subsection{Validation of the simulation approach}

The results of the combination of TCAD and Allpix Squared were compared to the results from a TCAD-alone approach. Identical simulation conditions were used, including mobility and recombination models. To emulate a transition of charged particles but neglecting contributions from Landau fluctuations, a vertical line of charge carriers with a density of 63 eh/\si{\micro\meter} is injected through a \SI{10}{\micro\meter} thick epitaxial layer at the corner between four adjacent pixels. The induced current at the collection electrode is then compared between the two simulation approaches.

\begin{figure} [h]
\begin{center}
\begin{subfigure}{.5\textwidth}
  \centering
  \includegraphics[width=.85\linewidth]{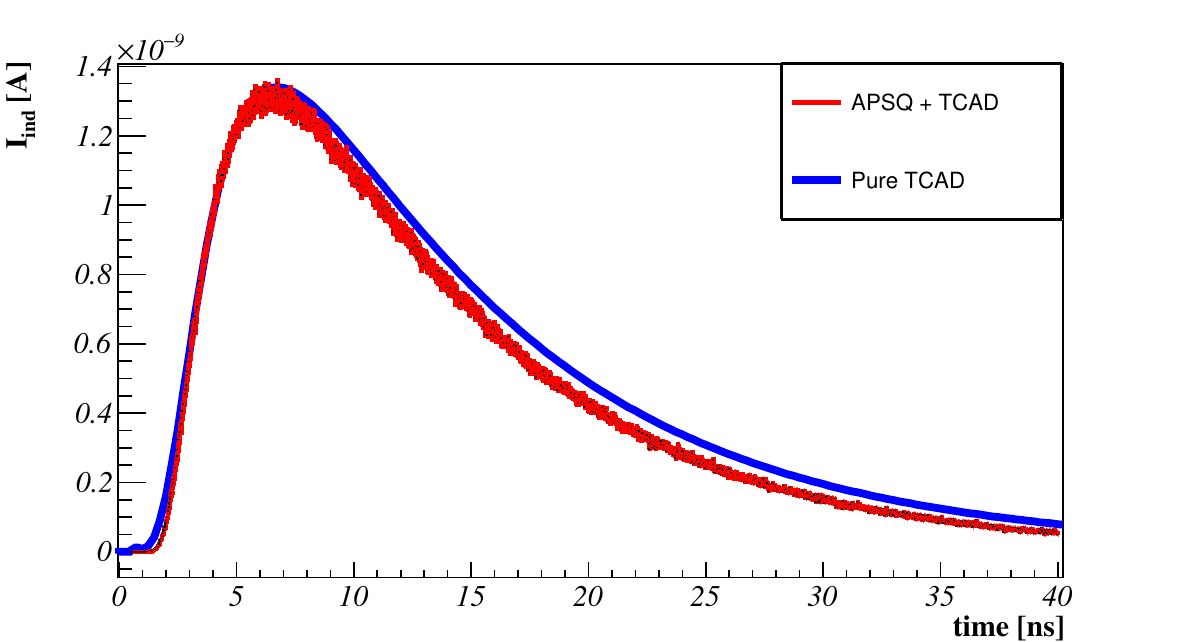}
  \caption{Standard}
  \label{fig:Standard_validation}
\end{subfigure}
\begin{subfigure}{.5\textwidth}
  \centering
  \includegraphics[width=.9\textwidth]{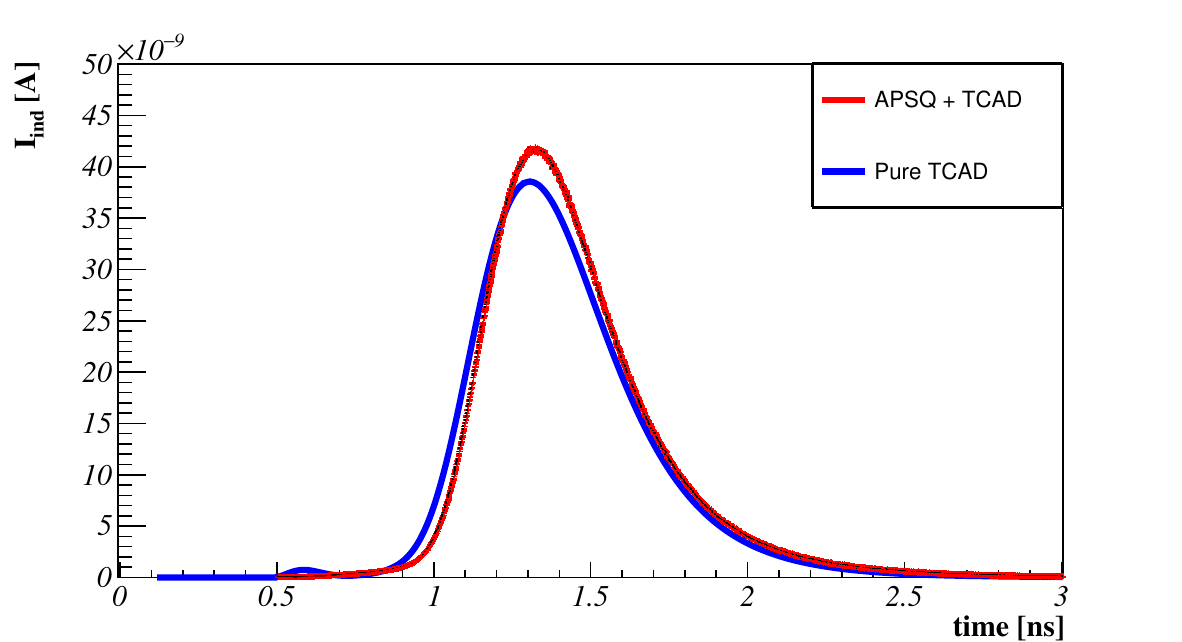}
  \caption{N-Gap}
  \label{fig:Ngap_validation}
\end{subfigure}
\caption{Transient simulation using TCAD and TCAD + APSQ for corner injection. Note the difference in time and current scales between the layouts.}
\label{fig:Layouts_validation}
\end{center}
\end{figure}

The results of the validation are shown in Figure \ref{fig:Layouts_validation}. A single pulse comprises the pure TCAD output, while an average of 1000 pulses is shown for the combined approach. A good qualitative agreement between the approaches is achieved for both layouts, while small deviations can be observed, most likely arising from differences in the calculation of the induced current between frameworks. However, a similar value for the total charge collected was obtained by integrating the pulses over time. 

Studies of the sensor layout and how it affects the charge collection were performed by charge injection at different pixel positions. In particular, charge injection in the corner of the pixel will lead to maximum charge sharing and diffusion. The charge collection time is defined as the time required for the signal to collect ninety-five percent of the total charge. A slow charge collection of around \SI{40}{\nano\second} can be seen in Figure \ref{fig:Standard_validation}. This is due to the small depleted region confined around the collection electrode. Larger currents and a collection time of around \SI{3}{\nano\second} were obtained on the N-Gap layout shown in Figure \ref{fig:Ngap_validation}, which highlights the timing capabilities of this layout.

\subsection{Simulations with Minimum Ionizing Particles}
The next step is to proceed to simulate the traversal of minimum ionizing particles and thus take into account Landau fluctuations, secondary particle production, and photo-ionization including contributions from the substrate. In the following, only the results of the N-Gap layout will be discussed.

\begin{figure} [h]
\begin{center}
\includegraphics[width=.55\linewidth]{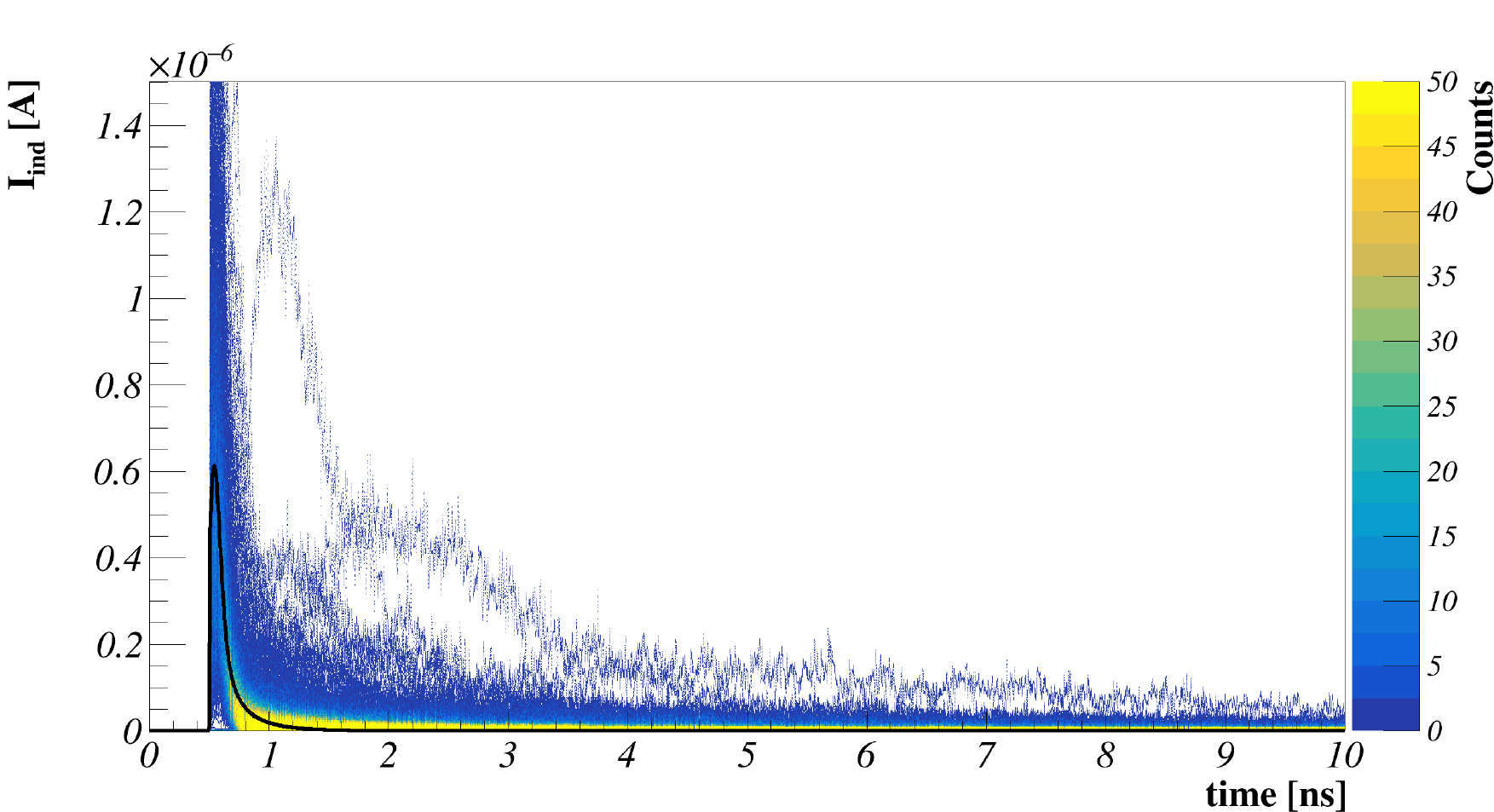}
\caption{Pulses induced in a \SI{50}{\micro\meter} thick N-Gap sensor using \SI{4}{\giga\electronvolt} electrons at the center of the collection electrode. In black, the average pulse obtained using 63 eh/\si{\micro\meter} injected charge is shown for comparison.}
\label{fig:Landau_Ngap}
\end{center}
\end{figure}

\begin{figure}[h]
\begin{center}
\includegraphics[width=.55\linewidth]{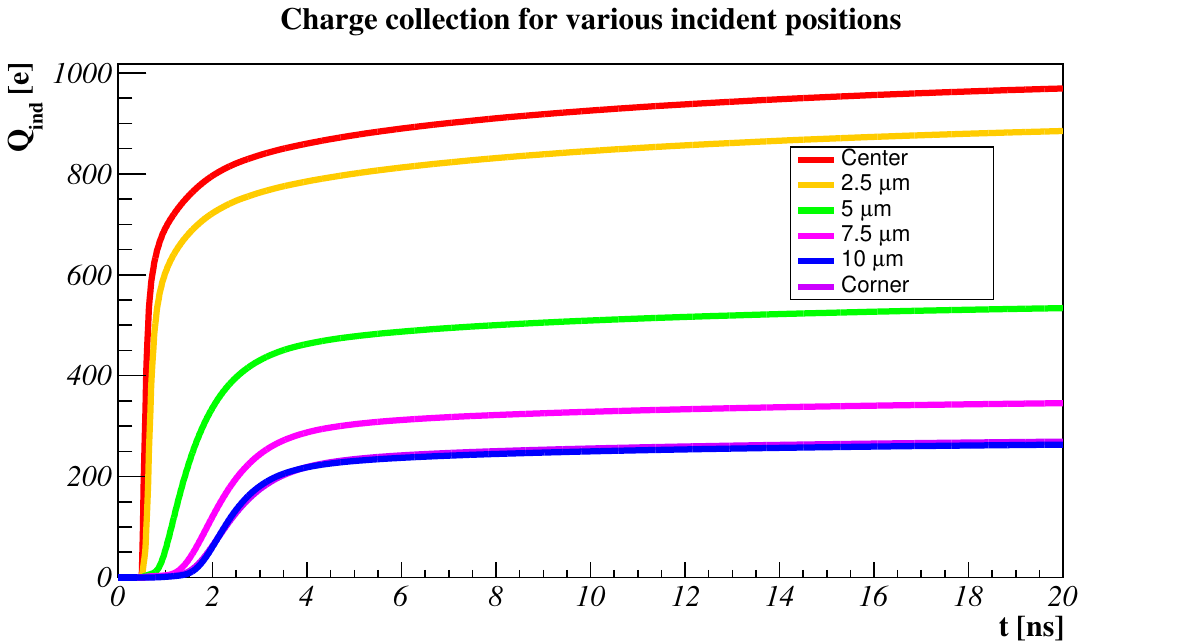}
\caption{Average charge collected at various incident positions along the pixel diagonal of the N-Gap layout. A larger amount of collected charge is obtained near the collection electrode.}
\label{fig:Charge_Collection}
\end{center}
\end{figure}

An overlay of the resulting induced pulses in comparison with the TCAD response using linear energy transfer is shown in Figure \ref{fig:Landau_Ngap}. Landau fluctuations are visible. These event-by-event fluctuations in energy deposition lead to a varying amount of charge carriers during ionization. A thicker substrate also contributes to a higher charge average. 

To test how this affects the average charge collection, a simulation using a \SI{4}{\giga\electronvolt} electron beam at different positions along the pixel diagonal was performed. By integration of the induced current signal, the collected charge of each event can be calculated. The results of the average charge collection resulting from 1000 primary particles at each incident position are shown in Figure \ref{fig:Charge_Collection}. A larger amount of collected charge is obtained for particles traversing near the center at the collection electrode while due to charge sharing and a reduced electric field strength, less charge is collected for particles traversing near the pixel corner. For a threshold value of around 160 electrons, an expected value of the working threshold for these sensors \cite{article}, an average charge collection of less than \SI{4}{\nano\second}  is achieved regardless of the position, which shows a promising timing capability of the N-Gap layout.    

\begin{comment}
\section{Test Beam Setup}
To investigate the actual capabilities of the sensor, a test beam campaign at the DESY-II facility \cite{DesyII} was performed in May of 2023. A Standard, 25x\SI{25}{\square\micro\meter} APTS with an epitaxial and substrate thickness of \SI{50}{\micro\meter} was put in the middle of a EUDET-type telescope \cite{EUDET}, consisting of six MIMOSA-26 planes \cite{MIMOSA}. A Telepix \cite{Telepix} sensor was used in coincidence with a scintillator as a trigger and as a time reference as well. The four central pixels of the APTS were connected to an oscilloscope to record waveforms related to a triggered event for offline analysis. 

The objective of the test beam was to reconstruct tracks in the DUT and relate waveforms to each track. With this information, can be studied how the incident position and charge fluctuations affect the rise time. The rise time is an intrinsic quantity of the signal development and the front-end circuitry which can be compared with simulation results using the transient simulation output and the implementation of front-end simulations of the chip. 

\end{comment}

\section{Conclusions and Outlook}
A combination of TCAD and Allpix Squared simulations has been compared with results from a TCAD-alone approach. A good qualitative agreement was achieved for both layouts. The combination of TCAD and Allpix Squared allowed to include stochastic effects such as Landau fluctuations and produced more realistic simulation scenarios that can be later compared to data obtained during test beam campaigns. Simulation studies on charge collection at different incident positions were performed to understand the timing capabilities of both layouts. The N-Gap layout achieves a collection time shorter than \SI{4}{\nano\second} regardless of the incident position for a threshold of around 160 electrons. This result is promising for the timing capabilities of this layout due to its charge collection time and consistency.

A test beam campaign was performed at the DESY-II facility \cite{DESYII} to obtain data to compare with simulations and calibration between charge and the electronic output in voltage units that was necessary to achieve this. The latter was performed with the use of an Iron-55 source due to its precise and relatively low energy peaks compared to other available sources and the linearity between charge and voltage observed with test pulses using a charge injection capacitor. The reliability of these results and an agreement between simulations and data will allow to obtain future predictions on different prototype designs.

\section{Acknowledgments}
The TANGERINE project is funded by the Helmholtz Innovation Pool,2021–2024, Germany. This work has been sponsored by the Wolfgang Gentner Programme of the German Federal Ministry of Education and Research, Germany (grant no. 13E18CHA) and has received funding from the European Union’s Horizon 2020 Research and Innovation programme (GA no. 101004761). The measurements leading to the data that will be used to compare these results have been performed at the Test Beam Facility at DESY Hamburg (Germany), a member of the Helmholtz Association (HGF).

The authors express their gratitude to the CERN EP R\&D WP 1.2, the APTS designers, and the ALICE ITS3 measurement team for their support.

\bibliographystyle{unsrt} % We choose the "plain" reference style
\bibliography{references}

\end{document}